\documentclass[twocolumn,aps,showpacs,prc]{revtex4}
\usepackage{graphicx}
\usepackage[dvips]{epsfig}
\usepackage{amsmath,bm}
\begin{document}
%\draft

\title{Gravitational waves from non-radial perturbations in glitching pulsars}

\author{ Joydev Lahiri$^{1}$ and D. N. Basu$^{2}$ }

\affiliation{Variable  Energy  Cyclotron Centre, 1/AF Bidhan Nagar, Kolkata 700064, India}

\email[E-mail 1: ]{joy@vecc.gov.in}
\email[E-mail 2: ]{dnb@vecc.gov.in}

\date{\today }

\begin{abstract}

    The Rossby mode (r-mode) perturbations in pulsars as a steady gravitational wave (GW) sources have been explored. The time evolution and the intensity of the emitted GWs in terms of the strain tensor amplitude have been estimated with the approximation of slow rotation adopting the equation of state derived using the Skyrme effective interaction with NRAPR parameter set. The core of the neutron star has been considered to be $\beta$-equilibrated nuclear matter composed of neutrons, protons, electrons and muons, which is surrounded by a solid crust. Calculations have been made for the critical frequencies, the evolution of frequencies and frequency change rates with time as well as the fiducial viscous and gravitational timescales, across a broad spectrum of pulsar masses. Our findings reveal that the r-mode instability region is associated with rotating young and hot pulsars. Furthermore, it is noteworthy that pulsars with low $L$ value emit gravitational radiation and fall within the r-mode instability region if the primary dissipative mechanism is shear viscosity along the crust-core interface boundary layer. The r-mode perturbation amplitude increases because of GW emissions, in contrast to other non-radial perturbations which transport to infinity the star's angular momentum. Thus the presence of these stellar perturbations implies a non-negative rate of change in transfer of rotational angular momentum. This observation suggests that for a glitching pulsar, the GW emission intensity evolves increasingly with time till the angular frequency diminishes to a value that is below a crucial threshold, after which the compact star ceases to emit radiation. 
\vspace{0.2cm}    

\end{abstract}


\maketitle

\noindent
\section{Introduction}
\label{Section 1} 
     
    The rapid rotations of accreting isolated neutron star (NS) which generate quasi-normal modes act as a sensitive probe for the effects of general relativity such as gravitational waves (GWs) and the properties of ultradense matter. The variations in the period of rotation of compact stars with time can reveal their internal fluctuations such as the core-crust coupling and decoupling. The GWs from glitching pulsars due to their rotational instabilities can explore the high density behavior of an Equation of State (EoS). The experimental detections of these GWs by LIGO and VIRGO, recently, provide new vistas in asteroseismology.  
    
    The instabilities in rotation of compact stars appear in distinct flavors, but they have one common feature in general. That is, these can directly be correlated to the unstable modes \cite{Andersson1998,Andersson2003,Freidman1998,Provost1981,Andersson2001,Bondarescu2007} of oscillations. Here the r-mode instability has been studied in the context of NSs using EoS obtained from the effective nucleon-nucleon (NN) interaction described by Skyrme (NRAPR) parameter set. Anderson's \cite{Andersson1998} numerical explorations of r-mode oscillation in NSs, supplemented by Friedman and Morsink's \cite{Freidman1998} theoretical work, have provided novel pathways for investigating GWs released by compact stars using cutting-edge detection technology. The detection of GWs may offer potential insights into the spin up of cold accreting NSs as well as the spin down mechanism observed in hot young compact stars.
 
    The r-mode oscillation is caused by perturbations in the velocity field of a star due to small disturbances in its density. Whereas the Coriolis effects provide a weak restoring force in a rotating star, in a non-rotating star the r-modes are neutral rotational motions. The frequency of r-mode always appears with reversed signs in the rotating and inertial frames. This implies that when observed from the inertial frame, an observer will perceive the modes as prograde while they will appear retrograde in the rotating system. The frequency of the mode, to the leading order, is given by \cite{Andersson1999,Papaloizou1978} 
    
\begin{equation}
\sigma=\frac{\left(l-1\right)\left(l+2\right)}{l\left(l+1\right)} \Omega
\label{eq1}
\end{equation}

    As $0<\sigma<\Omega$ $\forall$ $l\geq 2$, with $\Omega$ being the star's angular velocity in the inertial frame, by Chandrasekhar-Friedman-Schutz (CFS) mechanism the r-modes are destabilized and are unstable due to the GW emissions. The radiation of GWs resulting from r-modes originates from mass currents that vary over time, making it a gravitational equivalent to the magnetic monopole radiation. In NSs, compared to any other mode, the instability due to gravitational radiation for $l=2$ quadrupole r-mode is stronger. Further, these modes exist with velocity perturbation iff $l=m$ \cite{Provost1981, Andersson1999}. In spite of a decrease in energy $E_{inl}$ in the inertial-frame, the GW emissions cause increase in the energy $E_{rot}$ of the mode in the rotating frame. This baffling effect can be understood from the following relation,
    
\begin{equation}
E_{rot}=E_{inl}-\Omega J
\label{eq2}
\end{equation}
with $J$ being the star's angular momentum. It is clear from above that if both $J$ and $E_{inl}$ decrease, it is quite possible that $E_{rot}$ may increase. Expressing in terms of expansion in angular velocity $\Omega$, these r-mode frequencies, in leading order is \cite{Papaloizou1978,Lindblom1998}

\begin{equation}
\omega=-\frac{\left(l-1\right)\left(l+2\right)}{l+1} \Omega.
\label{eq3}
\end{equation}

    As the viscosity \cite{Lindblom1987} opposes the GW emission, the instability in the mode grows. To ensure the instability to be pertinent, it should increase faster than the viscous dampening. Thus the time scale for gravitation driven instability should be sufficiently short compared to time scale of viscous damping . The time dependence pf r-mode amplitude evolves as $ e^{i\omega t - t/\tau }$ due to the joint influence ordinary hydrodynamics and various dissipative processes. The effects of viscosity, gravitational radiation, etc. \cite{Lindblom2000,Lindblom1998,Owen1998} dictate the imaginary part of the frequency $1/\tau$. The associated time-scales of different processes involve the actual physical parameters of NS. While computing these physical parameters, nuclear physics plays a significant role in limiting the uncertainties in nuclear EoS. However, our current understanding of nuclear EoS is highly uncertain when it comes to high isospin asymmetric dense matter. Hence correlating the r-mode observables to the EoS predictions, which were obtained by systematically varying the physical properties, helps constraining the uncertainties associated with EoS.
		      
\noindent
\section{Non-radial perturbations in rotating neutron stars} 
\label{Section 2}
 
    The competition between the gravitational radiation and the dissipative influence of viscosity results in the evolution of the r-modes. Therefore the present work necessarily concerns the study of the effects of radiation upon the energy evolution of the modes. Aforementioned phenomenon can be described through the following integral
    
\begin{equation}
\widetilde{E}=\frac{1}{2}\int{\left[ \varrho \delta \vec{v}.\delta \vec{v}^{*}+\left(\frac{\delta p}{\varrho}-\delta \Phi \right)\delta \varrho^{*}\right]}d^{3}r,
\label{eq4}
\end{equation}
for the fluid perturbation \cite{Lindblom1998,Lindblom1999}, which contains the mass density profile $\varrho(r)$ of the star. The quantities $\delta \vec{v}$, $\delta p$, $\delta \Phi$ and $\delta \varrho $ represent the velocity, the pressure, the gravitational potential and the density perturbations, respectively, caused by oscillation of the mode. The r-mode dissipative time scale can be expressed by \cite{Lindblom1998},

\begin{equation}
\frac{1}{\tau_{i}}=-\frac{1}{2\widetilde{E}}\left(\frac{d\widetilde{E}}{dt}\right)_{i},
\label{eq5}
\end{equation}
where the subscript '$i$' has been used for indexing the various dissipative mechanisms like viscous drags (shear and bulk) and GW emissions.

    The expression of the energy appearing in Eq.(4) for the mode can be simplified using the lowest order expression of r-mode $\delta \vec{v}$ and $\delta \rho$ \cite{Lindblom1998,Vidana2012} to the following one-dimensional integral 

\begin{equation}
\widetilde{E}=\frac{1}{2}\alpha_r^{2} R^{-2l+2} \Omega^{2} \int^{R}_{0} \varrho(r) r^{2l+2} dr, 
\label{eq6}
\end{equation}
where R represents the NS radius, the amplitude of the mode is given by the dimensionless quantity $\alpha_r$, $\Omega$ represents angular velocity of rotation of NS and the radial dependence of the mass density of NS is given by $\varrho(r)$. 

\subsection{Dissipative time scales in glitching pulsars}

    As the expression for $(\frac{d\widetilde{E}}{dt})$ arising because of gravitational radiation \cite{Thorne1980,Owen1998} and viscous drags \cite {Lindblom1991,Owen1998,Lindblom2000} are rather well known, evaluation of the imaginary part $\frac{1}{\tau}$ can be facilitated using Eq.(5). It is advantageous to break up $\frac{1}{\tau}$ as
\begin{equation}
\frac{1}{\tau(\Omega,T)}=\frac{1}{\tau_{GR}(\Omega,T)}+\frac{1}{\tau_{SV}(\Omega,T)}+\frac{1}{\tau_{BV}(\Omega,T)},
\label{eq7}
\end{equation}
where $1/\tau_{GR}$, $1/\tau_{SV}$ and $1/\tau_{BV}$ are, respectively, the contributions of gravitational radiation, shear viscosity and bulk viscosity which can be given by \cite{Owen1998,Lindblom2000}

\begin{eqnarray}
\frac{1}{\tau_{GR}}=-\frac{32 \pi G \Omega^{2l+2}}{c^{2l+3}} \frac{(l-1)^{2l}}{[(2l+1)!!]^2}\left(\frac{l+2}{l+1}\right)^{(2l+2)}\nonumber \\
 \times\int^{R_{c}}_{0}\varrho(r)r^{2l+2} dr, 
\label{eq8}
\end{eqnarray}
%
%\begin{equation}
%\frac{1}{\tau_{BV}}\approx \frac{4 R^{2l-2}}{(l+1)^2} \int {\xi |{\frac{\delta \rho}{\rho}}|^{2}} %d^{3}r\left(\int^{R_{c}}_{0}\rho(r)r^{2l+2} dr\right)^{-1},
%\label{eq9}
%\end{equation}
%
\begin{eqnarray}
\frac{1}{\tau_{SV}}=\left[\frac{1}{2\Omega} \frac{2^{l+3/2}(l+1)!}{l(2l+1)!!I_{l}}\sqrt{\frac{2\Omega R_{c}^{2} \varrho_{c}}{\eta_c}}\right]^{-1}\nonumber \\
\times\left[\int^{R_{c}}_{0} \frac{\varrho(r)}{\varrho_{c}}\left(\frac{r}{R_{c}}\right)^{2l+2} \frac{dr}{R_c}\right]^{-1}, 
\label{eq9}
\end{eqnarray}
where $G$ is the gravitational constant, $c$ is the velocity of light; $R_{c}$, $\varrho_{c}$, $\eta_{c}$ are, respectively, at the outer edge of the core the radius, the density and the shear viscosity of the fluid.

    The time scale of shear viscosity described in Eq.(9) was determined by examining the dissipation due to shear viscosity within the boundary layer separating the solid crust and liquid core. This calculation assumes that the crust is inflexible and hence stationary within a rotating frame \cite{Lindblom2000}.
		
		The crustal movement resulting from its mechanical involvement with the core causes $\tau_{SV}$ to rise by $(\frac{ \Delta v}{v})^{-2}$. The factor $(\frac{ \Delta v}{v})^{-2}$ is calculated by dividing the difference between the velocities of the outer edge of the core and the inner edge of the crust by the core's velocity \cite{Levin2001}.

    The impact of a solid crust on r-mode instability was initially calculated by Bildsten and Ushomirsky \cite{Bildsten2000}. Their findings revealed that the shear dissipation within the viscous boundary layer resulted in reduction of the viscous damping time scale exceeding $10^5$ in accreting old NSs and surpassing $10^7$ in young and hot rotating NSs. For $l=2$, the $I_{l}$ in Eq.(9) has a value of $I_{2}=0.80411$ \cite{Lindblom2000}.

    Moreover, for stellar temperatures below $10^{10}$ K the bulk viscous dissipation is rather insignificant and shear viscosity in this temperature range is the dominant dissipative mechanism. Since in this work the temperature range has been restricted to $T<10^{10}$ K, the shear dissipative mechanism alone has been considered. Present studies are similar to the one performed by Moustakidis \cite{Moustakidis2015}, where the primary investigation focused on the instability boundary and related quantities (e.g. critical angular frequency, temperature etc.) of a NS by using the Skyrme effective interaction with NRAPR parameter set \cite{St05,Ak98}, while considering the influence of both the equation of state (EoS) and gravitational mass.		
		
\subsection{The Rossby-mode instability in rotating pulsars}		

    The instability in the temperature range $T \leq 10^{10}$ K has been studied, where as mentioned above, the dominant dissipative mechanism at the boundary layer is the shear viscous drag and its time scale is given by Eq.(9) with $\eta_c$ being the viscous coefficient of the fluid. The dominant contribution to the shear, in the range of temperatures $T \geq 10^9$ K, is primarily from the neutron-neutron (nn) scattering and below $T \leq 10^9$, it is mainly the electron-electron (ee) scattering that contributes to shear \cite{Lindblom2000}. Therefore,
    
\begin{equation}
\frac{1}{\tau_{SV}}=\frac{1}{\tau_{ee}}+\frac{1}{\tau_{nn}},
\label{eq10}
\end{equation}
where $\tau_{ee}$ and $\tau_{nn}$ can be evaluated from Eq.(9) by using the corresponding values of $\eta_{SV}^{ee}$ and $\eta_{SV}^{nn}$ given by \cite{Flowers1979,Cutler1987}

\begin{equation}
\eta_{SV}^{ee}=6 \times 10^{6} \varrho^{2} T^{-2} ~~~~~({\rm g/cm/s}), 
\label{eq11}
\end{equation}

\begin{equation}
\eta_{SV}^{nn}=347 \varrho^{9/4} T^{-2} ~~~~~({\rm g/cm/s}), 
\label{eq12}
\end{equation}
where CGS system of units has been used for all the above quantities while the temperature T is in Kelvin. It is often useful to factorize $\tau_{SV}$ and $\tau_{GR}$ by defining fiducial time scales $\widetilde{\tau}_{SV}$ and $\widetilde{\tau}_{GR}$ which facilitate transparent visualization of the roles of temperature and the angular velocity on different time scales. Hence, the fiducial shear viscous time scale $\widetilde{\tau}_{SV}$ can be defined as \cite{Lindblom2000,Lindblom1998},

 \begin{equation}
\tau_{SV}=\widetilde{\tau}_{SV} \left(\frac{\Omega_0}{\Omega}\right)^{1/2} \left(\frac{T}{10^8 K}\right),
\label{eq13}
\end{equation}
while the fiducial time scale for gravitational radiation $\widetilde{\tau}_{GR}$ can be defined as,
 
\begin{equation}
\tau_{GR}=\widetilde{\tau}_{GR} \left(\frac{\Omega_0}{\Omega}\right)^{2l+2},
\label{eq14}
\end{equation}
where $\Omega_0=\sqrt{ \pi G \bar{\rho}}$ and $\bar{\rho}= 3M/4 \pi R^3$ is the mean density of NS of radius $R$ and mass $M$. Therefore, neglecting bulk viscosity contributions, Eq.(7) reduces to

\begin{equation}
\frac{1}{\tau(\Omega,T)}=\frac{1}{\widetilde{\tau}_{GR}}\left(\frac{\Omega}{\Omega_0}\right)^{2l+2}+\frac{1}{\widetilde{\tau}_{SV}}\left(\frac{\Omega}{\Omega_0}\right)^{1/2} \left(\frac{10^8 K}{T}\right).
\label{eq15}
\end{equation}

    Dissipative effects cause the mode to decay exponentially as $e^{-t/\tau}$ i.e. the mode is stable, as long as $\tau>0$. It is evident from Eqns.(8) and (9) that $\widetilde{\tau}{SV}$ is positive while $\widetilde{\tau}{GR}$ is negative. As a result, viscosity has a stabilizing effect while gravitational radiation pushes these modes towards instability. Since the gravitational radiation is proportional to $\Omega^{2l+2}$, its impact on $1/\tau$ is very low for small $\Omega$. It implies that viscosity will dominate and the mode will be stable for sufficiently small angular velocities. On the contrary, for large values of $\Omega$ the gravitational radiation dominates and drives the mode towards instability. For a particular mode $l$ and a given temperature $T$, the condition $\frac{1}{\tau(\Omega_c,T)}=0$ defines the critical angular velocity $\Omega_c$. As a consequence, the equation of critical velocity for a particular mode with index $l$ and temperature $T$ can be expressed as a polynomial of degree $l+1$ in $\Omega_c^{2}$, resulting in a unique value of $\Omega_c$ for each mode. The $l=2$ being the smallest of these contributes the most, and hence studies have been done only for this mode in the present work. For the $l=2$ mode, the critical angular velocity $\Omega_c$ can be given by
    
\begin{equation}
\left(\frac{\Omega_c}{\Omega_0}\right)=\left(-\frac{\widetilde{\tau}_{GR}}{\widetilde{\tau}_{SV}}\right)^{2/11}\left(\frac{10^8 K}{T}\right)^{2/11}.
\label{eq16}
\end{equation}

    The mass shedding limit restricts the angular velocity that a NS can attain as maximum. This maximum angular velocity $\Omega_K$, called the Kepler velocity is given approximately by$\Omega_K\approx\frac{2}{3}\Omega_0$. Therefore, there exists a critical temperature below which the viscous drag completely suppresses the gravitational radiation. The above mentioned critical temperature $T_c$ can be given by \cite{Lindblom2000}
    
\begin{equation}
\frac{T_c}{10^8 K}=\left(\frac{\Omega_0}{\Omega_c}\right)^{11/2} \left(-\frac{\tilde{\tau}_{GR}}{\widetilde{\tau}_{SV}}\right)\approx(3/2)^{11/2}\left(-\frac{\widetilde{\tau}_{GR}}{\widetilde{\tau}_{SV}}\right).
\label{eq17}
\end{equation}
From Eq.(13) and Eq.(14), the critical angular velocity $\Omega_c$ can now be expressed as

\begin{equation}
\left(\frac{\Omega_c}{\Omega_0}\right)=\frac{\Omega_K}{\Omega_0}\left( \frac{T_c}{T}\right)^{2/11}\approx(2/3)\left( \frac{T_c}{T}\right)^{2/11},
\label{eq18}
\end{equation}
in terms of $T_c$, the critical temperature. Hence, once the EoS of NS is ascertained, the calculations of essential physical quantities to characterize the r-mode instability can be carried out.

    Because of the fact that the angular momentum is radiated to infinity by the gravitational radiation, the angular velocity evolves with time. Its time evolution, as per the approach of Owen et al. \cite{Owen1998}, can be given by
    
\begin{equation}
\frac{d\Omega}{dt}=\frac{2\Omega}{\tau_{GR}}\frac{\alpha_r^2Q}{1-\alpha_r^2Q},
\label{eq19}
\end{equation}
where the dimensionless quantities $\alpha_r$ is the r-mode amplitude and $Q=3 \widetilde{J}/2 \widetilde{I}$ where

\begin{equation}
\widetilde{J}=\frac{1}{MR^4}\int^{R}_{0}\varrho(r)r^{6} dr
\label{eq20}
\end{equation}
and
\begin{equation}
\widetilde{I}=\frac{8\pi}{3MR^2}\int^{R}_{0}\varrho(r)r^{4} dr.
\label{eq21}
\end{equation}
The dimensionless quantity $\alpha_r$ can be obtained from 'thermal' equilibrium or 'spin' equilibrium or may be treated as free parameter with values ranging from $1$ to $10^{-8}$. The solution for the angular frequency $\Omega(t)$ can be obtained by solving the equation represented by Eq.(19) as

\begin{equation}
 \Omega(t)=\left(\Omega^{-6}_{in}-\textsl{C}t\right)^{-1/6},
\label{eq22}
\end{equation}
by considering that the heat extracted by the neutrino emission is same \cite{Bondarescu2009,Moustakidis2015} as that generated by the shear viscosity under ideal conditions, where 

\begin{equation}
 \textsl{C}=\frac{12\alpha_r^2Q}{\widetilde{\tau}_{GR}\left(1-\alpha_r^2Q\right)}\frac{1}{\Omega_0^6},
\label{eq23}
\end{equation}
and $\Omega_{in}$, whose value conforms to initial angular velocity, has been treated as a free parameter. Using Eq.(22) and Eq.(23), the spin down rate given by Eq.(19) reduces to

\begin{equation}
\frac{d\Omega}{dt}=\frac{\textsl{C}}{6}\left(\Omega^{-6}_{in}-\textsl{C}t\right)^{-7/6}.
\label{eq24}
\end{equation}

    The spin of a NS decreases continuously till it reaches $\Omega_c$. The time $t_c$ elapsed since a NS evolves from the initial angular velocity $\Omega_{in}$ to $\Omega_{c}$ which is its minimum value and can be given as
    
\begin{equation}
t_c=\frac{1}{\textsl{C}}\left(\Omega_{in}^{-6}-\Omega_{c}^{-6}\right).
\label{eq25}
\end{equation}

\noindent
\section{Equation of State for dense nuclear matter} 
\label{Section 3}
    
    T. H . R. Skyrme developed the Skyrme interaction in 1959 \cite{Sk56}. In the year 1972, Vautherin and Brink did Hartree-Fock (HF) calculations for spherical nuclei using Skyrme's density-dependent effective NN interaction. They found a remarkable description of doubly-closed shell nuclei ground-state properties \cite{Va72}. Since Skyrme's pioneering work, and the Brink and Vautherin parametrization of the original interaction, constant attempts have been made by various groups to modify the Skyrme-type effective NN interaction parameters and confine Skyrme interaction parameter sets in order to better recreate experimental results. The analytical simplicity of the interactions is a main advantage allows one to determine the parameters that include fundamental properties.    

\subsection{Skyrme interaction and equation of state}  
    
    The standard form of the Skyrme interaction \cite{Sk56,Ch97} is given by 
		
\begin{eqnarray}
V({\bf r_1},{\bf r_2})&=&t_0 (1+x_0 P_\sigma)\delta({\bf r})\\ \nonumber 
&+&\frac{1}{2}t_1(1+x_1 P_\sigma)[{\bf P}^{'2}\delta({\bf r})+\delta({\bf r}){\bf P}^{2}]\\ \nonumber
&+&t_2(1+x_2 P_\sigma)[{\bf P}^{'}\bm{\cdot}\delta({\bf r}){\bf P}]\\ \nonumber
&+&\frac{1}{6}t_3(1+x_3 P_\sigma)[\rho({\bf R})]^\gamma\delta({\bf r})\\ \nonumber
&+&iW_0\bm{\sigma}\bm{\cdot}[{\bf P}^{'}\times\delta({\bf r}){\bf P}]\\ \nonumber
\label{seqn26}
\end{eqnarray}
\noindent
where $\rho$ is the number density, ${\bf r}={\bf r_1}-{\bf r_2}$, ${\bf R}=\frac{1}{2}[{\bf r_1}+{\bf r_2}]$, ${\bf P}=\frac{1}{2i}[\bm{\nabla}_1-\bm{\nabla}_2]$, ${\bf P}^{'}$ is the complex conjugate of ${\bf P}$ acting on the left, $\bm{\sigma}=\bm{\sigma}_1+\bm{\sigma}_2$ and $P_\sigma=\frac{1+\bm{\sigma}_1\bm{\cdot}\bm{\sigma}_2}{2}$. In the above equation the first term represents central term, second and third terms represent non-local term, fourth term represents density dependent term and last term represents spin-orbit term. The Skyrme interaction thus collectively contains 9 parameters. These are $t_0$, $t_1$, $t_2$, $t_3$, $x_0$, $x_1$, $x_2$, $x_3$ and $\gamma$. Using density functional the energy per particle $\epsilon_b$ of an asymmetric infinite nuclear matter (ANM) can be represented as

\begin{figure}[ht!]
\vspace{0.0cm}
\eject\centerline{\epsfig{file=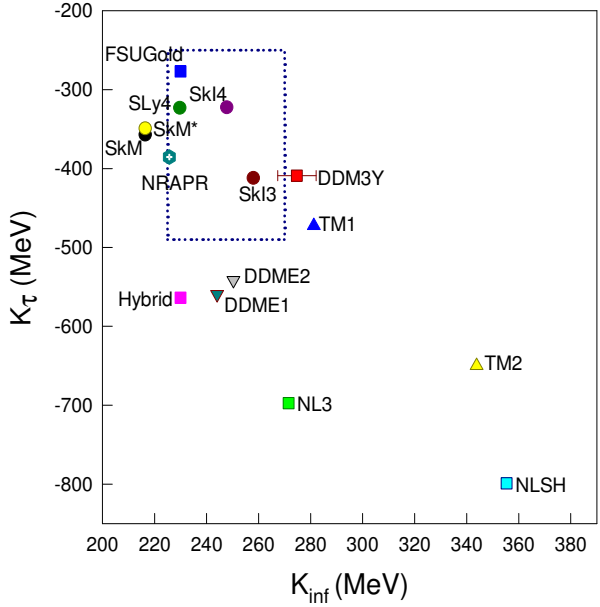,height=8cm,width=7.85cm}}
\caption{Plots of isobaric incompressibility K$_\tau$ versus SNM incompressibility K$_\infty$ (K$_{inf}$). Present results for NRAPR interaction is compared with other interactions \cite{Pi09,Sa07}. The rectangular area encloses K$_\infty$ values range 225-270 MeV \cite{Sh09} together with K$_\tau$ values equal to -370$\pm$120 MeV \cite{Chen09}.}
\label{fig1}
\vspace{0.0cm}
\end{figure}

\vspace{-0.4cm}
\begin{eqnarray}
\epsilon_b(\rho,\alpha)&=&\frac{3\hbar^2}{10m}\Big(\frac{3\pi^2}{2}\Big)^{2/3}\rho^{2/3}F_{5/3}\\ \nonumber
&+&\frac{1}{8}t_0\rho[2(x_0+2)-(2x_0+1)F_2]\\ \nonumber 
&+&\frac{1}{48}t_3\rho^{\gamma+1}[2(x_3+2)-(2x_3+1)F_2]\\ \nonumber
&+&\frac{3}{40}\Big(\frac{3\pi^2}{2}\Big)^{2/3}\rho^{5/3}[t_1(x_1+2)+t_2(x_2+2)]F_{5/3}\\ \nonumber
&+&\frac{3}{40}\Big(\frac{3\pi^2}{2}\Big)^{2/3}\rho^{5/3}\frac{1}{2}[t_2(2x_2+1)-t_1(2x_1+1)]F_{8/3}\\ \nonumber
\label{seqn27}
\end{eqnarray}
\noindent
where $\alpha=(N-Z)/A=1-2x_p$, $x_p=Z/A=\rho_p/\rho$ is the proton fraction, $F_m(x_p)=2^{m-1}[x_p^m+(1-x_p)^m]$ or $F_m(\alpha)=\frac{1}{2}[(1+\alpha)^m+(1-\alpha)^m]$.
		
    The pressure $P(\rho,\alpha)$ in ANM can be defined in terms of energy per particle or chemical potential as 

\begin{eqnarray}
P(\rho,\alpha)&=&\rho^2\frac{\partial\epsilon_b(\rho, x_p)}{\partial\rho}\\ \nonumber
&=&\mu_n\rho_n + \mu_p\rho_p - H(\rho,\alpha),\\ \nonumber
\label{seqn28}
\noindent
\end{eqnarray}
\noindent
where $H(\rho,\alpha)$ represents energy density of ANM and is related to energy per particle $\epsilon_b(\rho,\alpha)$ of ANM as $H(\rho,\alpha)=\rho\epsilon_b(\rho,\alpha)$ and
$\mu_{n(p)}~\Big(=\frac{\partial H(\rho,\alpha)}{\partial\rho_{n(p)}}\Big)$ is the neutron (proton) chemical potential. The pressure $P(\rho,\alpha)$ in ANM may then be expressed as

\begin{eqnarray}
P(\rho,\alpha)&=&\frac{\hbar^2}{5m}\Big(\frac{3\pi^2}{2}\Big)^{2/3}\rho^{5/3}F_{5/3}\\ \nonumber
&+&\frac{1}{8}t_0\rho^2[2(x_0+2)-(2x_0+1)F_2]\\ \nonumber 
&+&\frac{1}{48}t_3(\gamma+1)\rho^{\gamma+2}[2(x_3+2)-(2x_3+1)F_2]\\ \nonumber
&+&\frac{1}{8}\Big(\frac{3\pi^2}{2}\Big)^{2/3}\rho^{8/3}[t_1(x_1+2)+t_2(x_2+2)]F_{5/3}\\ \nonumber
&+&\frac{1}{8}\Big(\frac{3\pi^2}{2}\Big)^{2/3}\rho^{8/3}\frac{1}{2}[t_2(2x_2+1)-t_1(2x_1+1)]F_{8/3}\\ \nonumber
\label{seqn29}
\end{eqnarray}
\noindent

\begin{figure}[ht!]
\vspace{0.0cm}
\eject\centerline{\epsfig{file=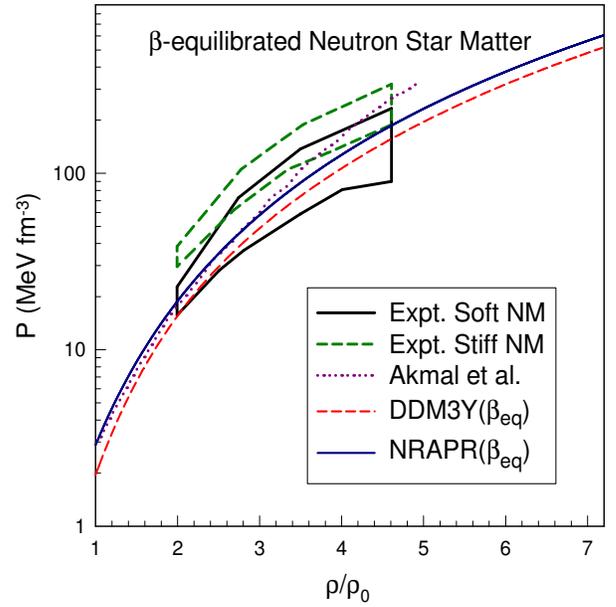,height=8cm,width=7.85cm}}
\caption{Plots of pressure $P$ versus $\rho/\rho_0$ for dense nuclear matter.}
\label{fig2}
\vspace{0.0cm}
\end{figure} 
    
    The outputs of 240 Skyrme interaction parameter sets, in 4 domains where experimentally or empirically derived 11 constraints exist, have been examined by Dutra et al. \cite{Du12}. These domains consist of a detailed systematic analyses of the symmetric nuclear matter (SNM) (4 sets named SM1, SM2, SM3, SM4 in Dutra et al.) properties, pure neutron matter (PNM) (2 sets named PNM1, PNM2 in Dutra et al.) properties and both the SNM and the PNM (5 sets named MIX1, MIX2, MIX3, MIX3 and MIX5 in Dutra et al.) properties. These domains have been covered by quite a few selected macroscopic constraints. It was observed that only six satisfy all the constraints out of the 240 Skyrme models whereas 66 satisfy all the properties except one. In 10 out of the 66, even those having only one failure have their magnitudes off by $\leq$ 5$\%$. The final list includes 16 consistent models (the CSkP set) consisting of GSkI, GSkII, KDE0v1, LNS, MSL0, NRAPR, Ska25s20, Ska35s20, SKRA, SkT1, SkT2, SkT3, Skxs20, SQMC650, SQMC700 and SV-sym32. These models satisfy a host of criteria extracted from the macroscopic properties of nuclear matter in the neighborhood of the nuclear saturation density at zero temperature and their density dependence obtained from the liquid drop model, experiments with giant resonances and heavy-ion collisions. Further curtailment in this number to 5 ensues with the application of the constraints like density dependence of the proton and the neutron effective masses, Landau parameters of SNM and PNM, $\beta$-equilibrated matter and observational data on low- and high-mass cold NSs. These five types of Skyrme models have been termed collectively as the CSkP$^*$ set consisting of KDE0v1, LNS, NRAPR, SKRA and SQMC700.

\begin{figure}[ht!]
\vspace{0.0cm}
\eject\centerline{\epsfig{file=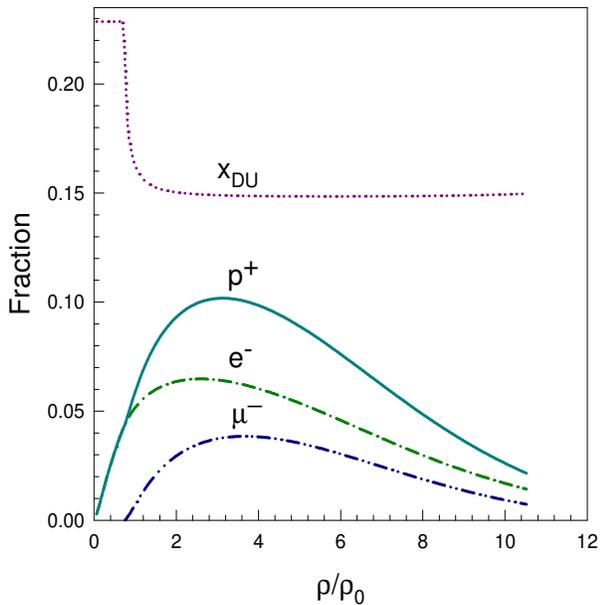,height=8cm,width=7.85cm}}
\caption
{The $\beta$-equilibrium electron ($e^-$), muon ($\mu^-$) and\\ proton ($p^+$) fractions plotted against the baryon density.\\ The Direct-URCA proton fraction threshold ${\rm x}_{\rm DU}$ is also\\ shown.} 
\label{fig3}
\vspace{0.0cm}
\end{figure}
    
    As extrapolation to densities above the regions of pressures, which show consistencies with the experimental flow data, is required for describing the NS structure, constraints like maximum mass and the corresponding central density of high-mass NSs put further restrictions on the Skyrme models. The radio pulsars, which are NSs with masses $>$ 1.8 $M_\odot$, are critical probes of nuclear astrophysics in extreme conditions. These massive NSs have extremely high gravitational fields inside, leading to substantially higher gravitational binding energies, than inside commonly found 1.4 $M_\odot$ NSs. The mass measurement sets an upper limit on this maximum density of 10 times the saturation density \cite{De10}. The maximum mass NS with central density in line with observation is not simulated by any of the CSkP$^*$ models except NRAPR and KDE0v1 parameter sets. For the present work, the EoS for $\beta$-equilibrated NS matter has been derived by using the Skyrme interaction with NRAPR parameter set \cite{St05} provided in Table-I.

\begin{figure}[ht!]
\vspace{0.0cm}
\eject\centerline{\epsfig{file=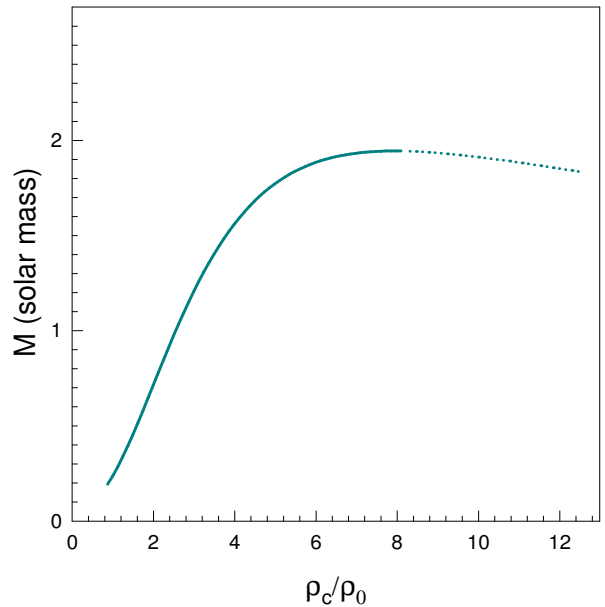,height=8cm,width=7.85cm}}
\caption
{Plot of slowly rotating neutron star mass versus central density obtained using the present nuclear Equation of State. The unstable branch of TOV solution is shown in part by the dotted line.} 
\label{fig4}
\vspace{0.0cm}
\end{figure}        

\begin{table*}
\vspace{0.0cm}
\centering
\caption{Values of the nine parameters of nuclear matter for the Skyrme interaction corresponding to NRAPR \cite{St05}}
\renewcommand{\tabcolsep}{0.05cm}
\renewcommand{\arraystretch}{1.2}
\begin{tabular}{|c|c|c|c|c|c|c|c|c|c|c|c|}\hline
\hline
$\gamma$ & $t_0$ & $x_0$ & $t_1$ & $x_1$ & $t_2$ & $x_2$ & $t_3$ & $x_3$\\\hline
0.14416&-2719.7&0.16154&417.64&-0.047986&-66.687&0.027170&15042& 0.13611 \\\hline
\multicolumn{9}{|c|}{Nuclear matter properties at saturation density} \\
\hline
\multicolumn{1}{|c|}{$\rho_0$ ($\mathrm{fm}^{-3}$)}&\multicolumn{1}{|c|}{$\epsilon_b(\rho_0)$ (MeV)} & \multicolumn{2}{c|}{$K_\infty (\rho_0)$ (MeV)}
& \multicolumn{1}{c|}{$K_\tau$ (MeV)} & \multicolumn{1}{c|}{$\frac{m^*}{m}(\rho_0,k_{f_0})$}
& \multicolumn{1}{c|}{$E_{sym} (\rho_0)$ (MeV)} & \multicolumn{2}{c|}{$L (\rho_0)$ (MeV)} \\
\hline
\multicolumn{1}{|c|}{0.1606}& \multicolumn{1}{|c|}{-15.86} & \multicolumn{2}{c|}{225.65} & \multicolumn{1}{c|}{-385.32}
& \multicolumn{1}{c|}{0.69} & \multicolumn{1}{c|}{32.79} & \multicolumn{2}{c|}{59.63} \\\hline
\end{tabular}
\vspace{0.0cm}
\end{table*} 
      		
	It is possible to experimentally ascertain the incompressibility of SNM by analyzing the compressional modes of nuclei, specifically through the isoscalar giant monopole resonance (ISGMR) and isoscalar giant dipole resonance (ISGDR). The inconsistencies found in the self-consistency of HF-RPA calculations \cite{Sh06} for the strength functions of ISGMR and ISGDR lead to shifts in the calculated centroid energies. These shifts may be larger than the experimental uncertainties in magnitudes. In fact, the low values of $K_\infty$ in the range of 210-220 MeV were predicted because of not using a completely self-consistent calculations with the Skyrme effective NN interactions \cite{Sh06}. When this drawback is corrected, the SLy4 type Skyrme parmetrizations predict $K_\infty \sim 230-240$ MeV \cite{Sh06}. The relativistic value of $K_\infty \sim 250-270$ MeV $\sim$  MeV is somewhat on the higher side. The experimental data of ISGMR, however, suggests $K_\infty$ to be $\approx$ 240 $\pm$ 20 MeV.
    
    The ISGDR data usually predict smaller values \cite{Lu04,Yo04} of $K_\infty$. The extraction of the value of $K_\infty$ is more problematic from ISGDR due to various reasons. In particular, at excitation energies \cite{Sh06} exceeding 30 MeV for $^{116}$Sn and 26 MeV for $^{208}$Pb, the maximum cross-section for ISGDR reduces sharply with high excitation energies. This decline is particularly steep and may bring the cross-section below the detection limits of current experimental techniques. The latest experimental values \cite{Yo05} for the incompressibility $K_\infty$ of SNM indicate that it is quite similar to the estimates of the relativistic mean field (RMF) model \cite{Pi09}. The results of microscopic calculations using Gogny effective interactions \cite{Bl80}, which include nuclei where pairing correlations are important, reproduce experimental data on heavy nuclei for which SNM incompressibility $K_\infty$ is approximately 220 MeV \cite{Bl95}. Hence, one can infer that the NRAPR value for $K_\infty=$ 225.7 MeV is a satisfactory theoretical outcome and falls within the suitable range of 225-270 MeV \cite{Chen09} for the incompressibility of SNM, which is also supported by previous research \cite{Vr03,Sh09}.

    The advanced quantum molecular dynamics transport model can be employed to simulate the collisions of $^{112}$Sn and $^{124}$Sn nuclei. These simulations reproduce isospin diffusion data from two different observables and the ratios of proton and neutron spectra. By comparing these data to calculations performed over a range of NSEs at $\rho_0$ and different representations of the density dependence of the NSE, the constraints on the density dependence of the nuclear symmetry energy (NSE) \cite{Kl06} at subnormal density can be determined \cite{Ts09}. With these constraints \cite{Ts09}, the results for $K_\infty$, $L$, $E_{sym}(\rho_0)$ and $K_\tau$ \cite{CBS09} for NRAPR are consistent. In Table-I, the values of SNM incompressibility $K_\infty$, symmetry energy $E_{sym}(\rho_0)$, slope of nuclear symmetry energy $L$ and isobaric incompressibility $K_\tau$ MeV are tabulated.

    The reasonable value of incompressibility still remains controversial \cite{Sh06}. Fig.-1 compares the present calculation (NRAPR) of $K_\tau$ versus $K_\infty$ with predictions from several other models such as SkI3, SkI4, SLy4, SkM, SkM*, DDM3Y, FSUGold, NL3, Hybrid \cite{Pi09}, NLSH, TM1, TM2, DDME1, and DDME2, as reported in Table I of Ref. \cite{Sa07}. The range of magnitude of current $K_\infty$ values, which fall between 225-270 MeV according to \cite{Sh09}, and $K_\tau$ values of $-370\pm120$ MeV from \cite{CBS09} are enclosed by a dotted rectangle. Although NRAPR, SkI3, SkI4, SLy4, DDM3Y \cite{BCS08} and FSUGold are inside the rectangular region, unlike NRAPR the magnitude of $L$ value obtained using SkI3 is 100.49 MeV that is substantially larger than the acceptable limit of $58.9\pm16$ MeV \cite{Wa09,Ag13,Li13,Ag17} while except for NRAPR and DDM3Y none can reach $\sim$ 2M$_\odot$. A different review \cite{Kl17} discovered that $E_{sym}(\rho_0)$ equals 31.7 $\pm$ 3.2 MeV and $L$ equals 58.7 $\pm$ 28.1 MeV, with $L$ having a notably greater error than $E_{sym}(\rho_0)$. Nonetheless, DDME2 is relatively near to the rectangular area for which $L$ is 51 MeV.	

    It should be noted that the incompressibility value of 271.76 MeV for SNM calculated by RMF-NL3 \cite{La97,La99} is significantly greater when compared to the value of 225.7 MeV obtained from the NRAPR model. In Fig.-2, the pressure $P$ of dense nuclear matter as functions of $\rho/\rho_0$ is shown. The PNM is depicted by the unbroken line, while the broken line represents $\beta$-equilibrated NS matter, and the dotted line represents the A18 model using the Akmal et al. \cite{Ak98} variational chain summation. The areas enclosed by the continuous and broken lines correspond to the pressure regions for neutron matter (NM), with soft NM and stiff NM corresponding to the pressures from asymmetry terms with weak and strong density dependence, respectively, consistent with experimental flow data \cite{Da02}. The present EoS obtained for NRAPR interaction parameters \cite{Du11} agrees well with the experimental flow data \cite{Da02} despite that these parameters have been tuned to reproduce the properties of finite nuclei. This justifies its extrapolation to high density since its high density behavior looks phenomenologically confirmed from Fig.-2.
		
\noindent     
\subsection{The $\beta$-equilibrated $npe\mu$ neutron star matter} 
\label{Section 4}
    
        The direct URCA process which has a significant influence on the nuclear EoS causes rapid cooling of NSs \cite{La91,Pe92}. In recent years, attention has been drawn to the direct URCA process in NSs which may be the primary mechanism for its rapid cooling. This can, however, occur only when the $\beta$-equilibrium proton fraction $x_p$ in the star is $\geq {\rm x}_{\rm DU}=1/9$, when only electrons are considered, and $\geq {\rm x}_{\rm DU}=1/[1+\{1+\frac{x_e^{1/3}+(1-x_e)^{1/3}}{2}\}^3]$, when both electrons and muons are considered where ${\rm x}_{\rm DU}$ is the direct URCA proton fraction threshold. At very high densities or relativistic energies the leptonic masses can be neglected compared to their kinetic energies and then electronic number density $\rho_e$ $\approx$ muonic number density $\rho_\mu$ implying $x_e=\rho_e/(\rho_e+\rho_\mu)\approx\frac{1}{2}$ and then $x_p\geq {\rm x}_{\rm DU}= 1/[1+(1+2^{-1/3})^3] \approx~0.148$. In case of the direct URCA process, the reactions which occur are the neutron decay $n \rightarrow p+e^-+\bar{\nu_e}$, electron capture $p+e^-\rightarrow n+ \nu_e$, neutron decay $n \rightarrow p+\mu^-+\bar{\nu_\mu}$ and muon capture $p+\mu^-\rightarrow n+ \nu_\mu$. It would be interesting to know whether present EoS favors or disfavors direct URCA process. In our study, the lepton energy per particle $\epsilon_L(\rho,x_p)$ is given by the relativistic, ideal Fermi-gas expression; in addition to $e^-$, $\mu^-$ is also considered as and when they are energetically favored. At $\beta$-equilibrium, one has ${\partial \over \partial x_p} \left ( \epsilon_b(\rho,x_p)+\epsilon_L(x_p) \right )=0$, where $\epsilon_b(\rho, x_p)$ is the baryonic energy  per particle including the rest masses. In Fig.-3, the $\beta$-equilibrium proton fraction thus obtained in the NS matter is displayed as a function of the baryon density invoking the condition of charge neutrality.
        
\begin{figure}[ht!]
\vspace{0.0cm}
\eject\centerline{\epsfig{file=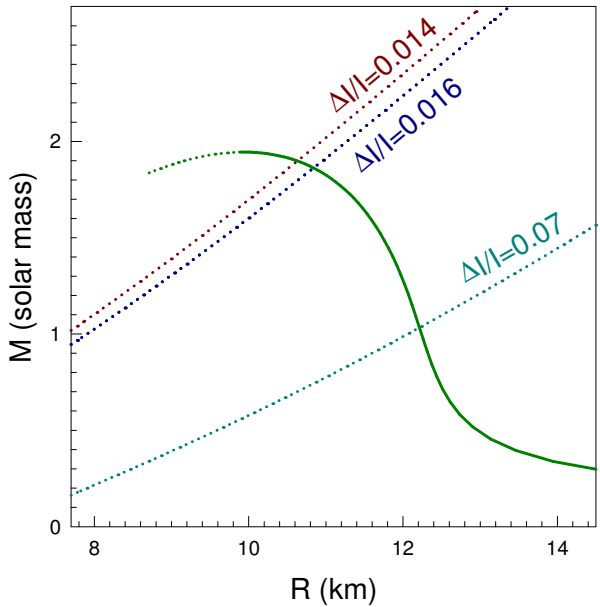,height=8cm,width=7.85cm}}
\caption{The mass versus radius plot for the slowly rotating\\ neutron stars. In the case of the Vela pulsar, if $\frac{\Delta I}{I} > 1.4\%$,\\ then the permissible values of the mass and the radius\\ are located on the right side of the curve which can be\\ described by the condition $\frac{\Delta I}{I} = 0.014$ (considering $\rho_t=$\\ 0.083 fm$^{-3}$ and the corresponding P$_t=$ 0.545 MeV fm$^{-3}$).\\ The unstable branch of TOV solution is shown in part\\ by the dotted line.} 
\label{fig5}
\vspace{-0.25cm}
\end{figure}

    Through the application of the Skyrme NRAPR effective NN interaction, the investigation of $\beta$-equilibrated dense matter stability in NSs has yielded the determination the inner edge location of their crusts and the density and pressure of their core-crust transition. The intrinsic stability, which pertains to the stability of any single phase, is maintained by the convexity of $\epsilon(\rho,x)$, as established by \cite{At14,La07}. 
		
    This requirement can be expressed thermodynamically as $V_{thermal}=\rho^2\Big[2\rho\frac{\partial \epsilon_b}{\partial \rho}+\rho^2\frac{\partial^2 \epsilon_b}{\partial \rho^2}-\rho^2\frac{(\frac{\partial^2 \epsilon_b}{{\partial\rho \partial x}})^2}{\frac{\partial^2 \epsilon_b}{\partial x^2}}\Big]$. By imposing $V_{thermal}=0$, the condition for core-crust transition is obtained. The incompressibility $K_\infty$ of SNM, the nuclear symmetry energy $E_{sym}(\rho_0)$ at saturation density $\rho_0$, the values for both the slope $L$ and the isospin-dependent component $K_\tau$ of the isobaric incompressibility have been presented in Table-I. These results align well with the constraints obtained from various measurements, such as the isotopic dependence of the giant monopole resonance in even-A Sn isotopes, the thickness of the neutron skin of nuclei, and the analysis of experimental data on isospin diffusion and isotopic scaling in heavy-ion collisions at intermediate energies.

\begin{figure}[ht!]
\vspace{0.0cm}
\eject\centerline{\epsfig{file=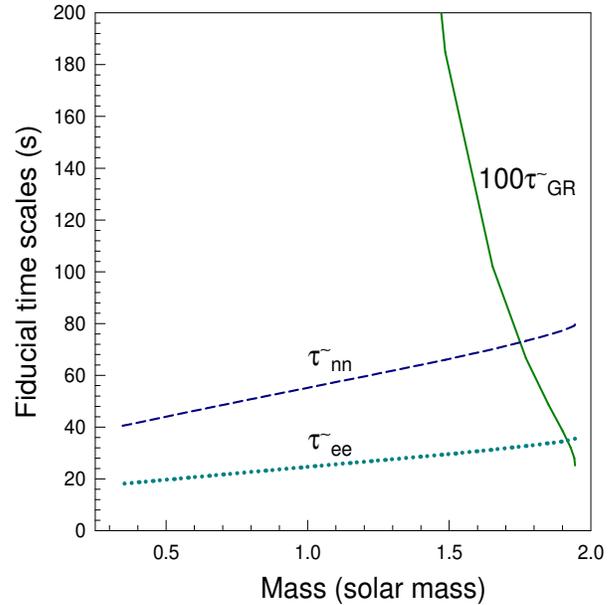,height=8cm,width=7.85cm}}
\caption{The fiducial timescales plotted as a function of neutron star gravitational masses obtained using NRAPR EoS.} 
\label{fig6}
\vspace{-0.2cm}
\end{figure}        
		
	The instability known as the r-mode can arise when the gravitational-radiation driving time scale of the mode is shorter than the dissipation time scales of various mechanisms present within a NS's interior. The nuclear EoS has two distinct impacts on the r-mode time scales. Firstly, it defines the radial distribution of mass density, which is the fundamental ingredient in the integrals that are relevant to the r-mode. Secondly, the EoS specifies the transition density from the neutron star core to its crust and the radius of the core, which act as the upper limits of these integrals. Using the NRAPR EoS, the Tolman-Oppenheimer-Volkoff equation has been solved to produce two plots, namely Fig.-4, which shows the relationship between mass and central density, and Fig.-5, which illustrates the mass-radius relationship of slowly rotating neutron stars. For a particular value of $\frac{\Delta I}{I}$, the mass-radius relation can be obtained using the density $\rho_t$ and pressure P$_t$ at the core-crust transition. Such mass-radius relations are then included in Fig.-5 for $\frac{\Delta I}{I} = 0.014, 0.016 ~{\rm and}~ 0.070$. The constraint $\frac{\Delta I}{I}>1.4\%$ for Vela pulsar, allows masses and corresponding radii to be to the right of the curve defined by $\frac{\Delta I}{I} = 0.014$ (for $\rho_t$ = 0.083 fm$^{-3}$ and P$_t$ = 0.545 MeV fm$^{-3}$) \cite{Du11}. On the basis of glitch activity of Vela pulsar, a recent observational data \cite{Ho12} suggests somewhat higher estimate for $\frac{\Delta I}{I}>1.6\%$. However, this small difference neither affects inferences drawn nor requires any new concept to extend partially the neutron superfluidity into the core. Nonetheless, when considering the occurrence of crustal entrainment resulting from the Bragg reflection of unbound neutrons by the lattice ions, the fraction of the moment of inertia attributed to the crust increases significantly (from 1.4-1.6$\%$ to 7$\%$) \cite{An12,Ch13}. The only reasonable limitation that can be put on Vela pulsar's mass is that it should exceed $\sim 1$ M$_\odot$ in conformity with the simulations of core-collapse supernova. This study, therefore, implies that the crust is not enough if entrainment is considered since the mass of Vela pulsar would be below 1 M$_\odot$ (Fig.-5), in accordance with other studies \cite{An12,Ch13,Ho15,De16,Do16}. However, the crust is enough to explain the data of Vela glitches in absence of entrainment. 
	
    A recent conjecture suggests a strong correlation between the slope of the symmetry energy $L$ and the density at which the core transits to the crust. This correlation is predicted to be independent of the relationship between $K_\tau$ and $L$ \cite{Du10}. However, there was no observed correlation between the transition pressure and $L$ \cite{Du10}. The maximum NS mass for the EoS obtained using the NRAPR Skyrme set is $\sim$ 1.9445 $M_\odot$ with a corresponding radius of $\sim$ 9.9267 kms where $M_\odot$ is the solar mass. It reconciles with the current observations on the massive compact stars $\sim$2 M$_\odot$ \cite{De10,An13}.    

\begin{figure}[ht!]
\vspace{0.0cm}
\eject\centerline{\epsfig{file=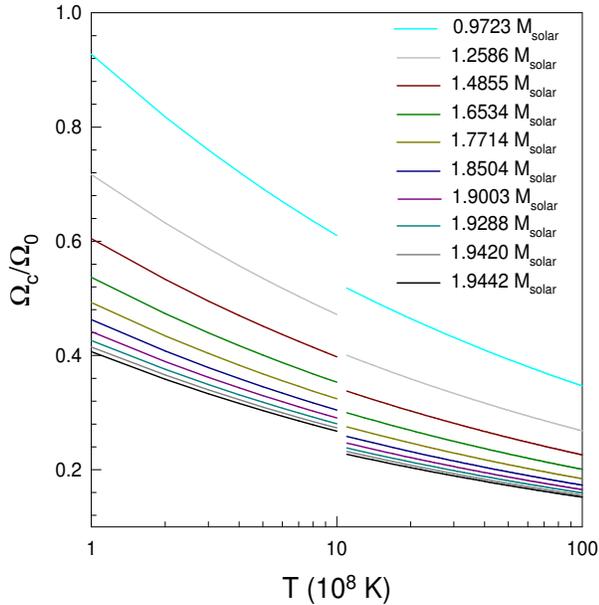,height=8cm,width=7.85cm}}
\caption{The reduced critical angular frequency plotted as a function of temperature for neutron stars of different gravitational masses.} 
\label{fig7}
\vspace{0.0cm}
\end{figure} 
    
\section{Gravitational waves from rotating compact stars}
         
\subsection{Constraints from Thermal Equilibrium}

    The gravitational radiation, in a steady-state, pumps energy into the r-mode at a rate given by \cite{Mahmoodifar2013}
    
\begin{equation}
  W_d =  (1/3)\Omega \dot J_c = -2\widetilde{E}/\tau_{GR}.
\label{eq30}
\end{equation} 
\noindent
This amount of energy, under thermal steady-state dissipates totally within the star. The amount of energy $L_\nu$ will be lost due to the neutrino emission and the rest due to emission of photons $L_\gamma$ \cite{Moustakidis2015,Bondarescu2009} both of which will get radiated from the star's surface. It is important to mention that when the mode becomes saturated \cite{Alford2014}, the thermal steady-state is independent of the cooling mechanism. Thus the thermal steady-state condition is $W_d=L_\nu+L_\gamma$ assuming that during quiescence all the emitted energy is entirely due to dissipation of r-mode within the star. The $J_c$ appearing in the Eq.(30) is the canonical angular momentum of the mode which is expressed as $J_c = -\frac{3}{2}\widetilde{J}M R^2 \Omega \alpha_r^2$, where $\widetilde{J}$ and $\widetilde{I}$ are the dimensionless quantities defined by Eqs.(20,21), respectively \cite{Mu18}. Using Eq.(5) and the explicit expression for $J_c$, the amplitude at saturation $\alpha_{th}$ in thermal equilibrium is then given by 

\begin{equation}
\alpha_{th} =\Big[\frac{-\tau_{GR}4\pi R^2\sigma T_{eff}^4}{\widetilde{J}M}\Big]^{1/2}\frac{1}{\Omega R}
\label{eq31}
\end{equation} 
\noindent
where $\sigma$ is the Stefan's constant.

\begin{figure}[ht!]
\vspace{0.0cm}
\eject\centerline{\epsfig{file=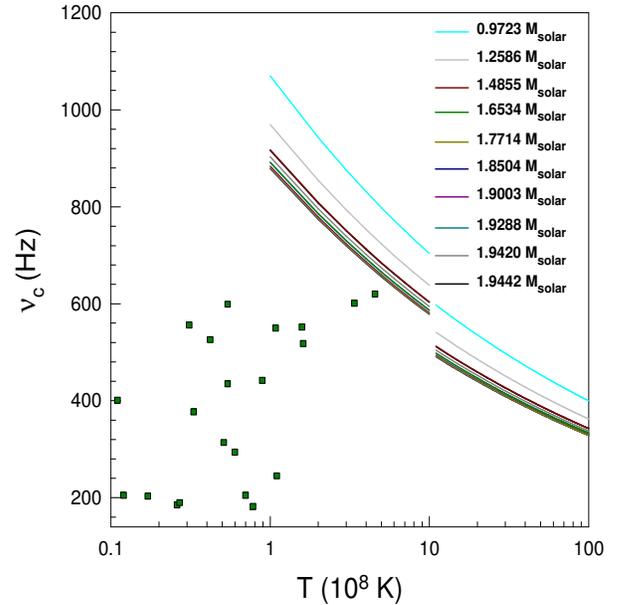,height=8cm,width=7.85cm}}
\caption{The critical angular frequency plotted as a function of temperature for neutron stars of different gravitational masses. The observational data \cite{Ha12} included in Table-IV are represented by square dots.} 
\label{fig8}
\vspace{0.0cm}
\end{figure} 
		
\subsection{Constraints from Spin Equilibrium}
    
    The NSs are sources of extremely high gravitational pull towards its center due to very high compactness. As a consequence, these stars can acquire mass from the companion if they belong to a binary system. Large angular momentum finally gets absorbed since any particle around the star can be free only at angular velocities reaching the Keplerian limit. While conserving the total angular momentum, the NS gains angular momentum by absorbing such particles of the companion star which is rotating with speed much lower than the Keplerian angular velocity. Therefore, a NS gradually spins up as it accumulates mass from its binary companion at a rate of $\dot M$. Hence the outburst properties and the observed spin-up rates may be used directly to constrain the amplitude of r-mode \cite{Brown2000, Ho2011}. Hence one has
    
\begin{equation}
  2\pi I \dot{\nu} \Delta =  \frac{2J_c}{\tau_{GR}}
\label{eq32}
\end{equation} 
\noindent
where $I=MR^2\widetilde{I}$ is the moment of inertia. The aforementioned equation displays the effect of r-mode perturbation on the right side, specifically the torque resulting from gravitational emission causing spin-down. The quantity $\dot{\nu}$ is the spin-up rate during outburst and $\Delta = (t_o/t_r)$ is the ratio of the outburst duration $t_o$ to the recurrence time $t_r$. Using Eq.(32) and the explicit expression for $J_c$, the amplitude at saturation $\alpha_{sp}$ in spin equilibrium follows:  

\begin{figure}[ht!]
\vspace{0.0cm}
\eject\centerline{\epsfig{file=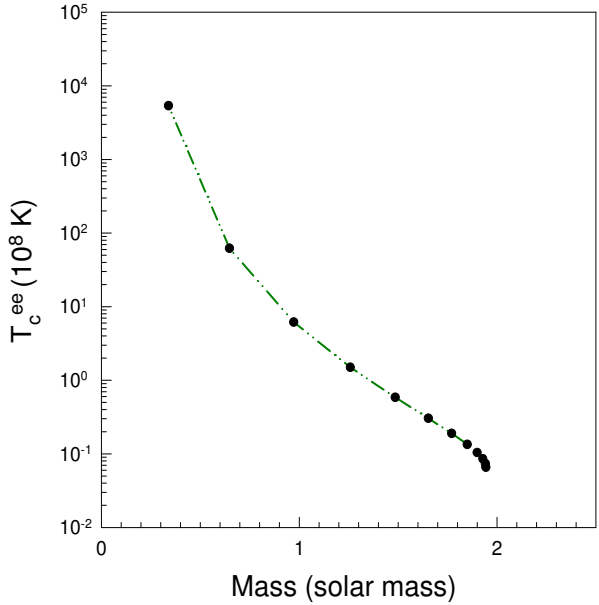,height=8cm,width=7.85cm}}
\caption{The critical temperature plotted as a function\\ of gravitational masses of neutron stars.} 
\label{fig9}
\vspace{0.0cm}
\end{figure}  
		
\begin{equation}
\alpha_{sp} =\Big[\frac{-\tau_{GR}2\pi \widetilde{I} \dot{\nu}\Delta}{3\widetilde{J}\Omega}\Big]
\label{eq33}
\end{equation} 
\noindent

\noindent     
\subsection{Gravitational Wave Amplitudes} 
 
     The angular momentum gets transferred as a NS accretes mass from its companion causing increase in its rate of rotation. Ultimately it surpasses $\Omega_c$, its critical value. The NS begins to emit GW at this epoch due to the perturbation of r-mode. The GW emission fuels this  perturbation due to r-mode because of CFS mechanism, resulting in an increase in the amplitude $\alpha_r$ until it saturates. As described earlier, its saturation value may be estimated either from the 'thermal equilibrium' or from the 'spin equilibrium'. The NS spins down to the region of stability by emitting GWs which take away the energy and the angular momentum with it. In the present calculations the emitted GW intensity radiated by NSs has been estimated and compared with the observational data of LIGO-Virgo collaboration (as Virgo was down during O1, only LIGO data was used) \cite{Abbott2019}. This has been described in terms of the strain tensor amplitude $h_0$ that is related to the amplitude $\alpha_r$ of the $r$-mode by \cite{Owen2009, Owen2010} as  
 
\begin{figure}[ht!]
\vspace{0.0cm}
\eject\centerline{\epsfig{file=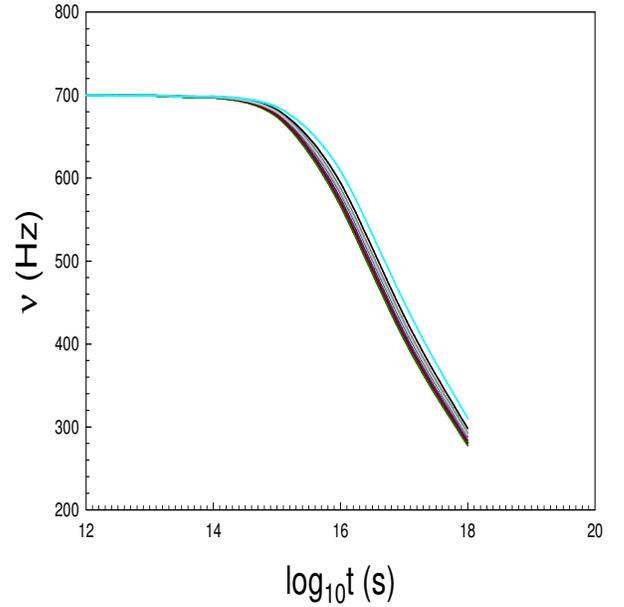,height=8cm,width=7.85cm}}
\caption{Plots of the evolving frequencies as a function of time.} 
\label{fig10}
\vspace{0.0cm}
\end{figure}
\vspace{0.0cm}  
      
\begin{eqnarray}
h_0=\sqrt{\frac{8\pi}{5}} \frac{G}{c^5} \frac{1}{r} \alpha_r \omega^3 M R^3\widetilde{J} ,
\label{eq34}
\end{eqnarray} 
\noindent
where the angular velocity of the star $\Omega$ is related to the r-mode angular frequency $\omega$ by the relation $\omega=-\frac{\left(l-1\right)\left(l+2\right)}{l+1} \Omega$. It is important to mention here that the signature of continuous emission of GW because of the perturbation due to r-mode is contradistinct from GW emission because of the ellipticity of a star as described in Ref.\cite{Owen2010}. In case of r-mode perturbation, the GW emission is commanded by the mass quadrupole moment while for ellipticity, it is dominated by the mass current quadrupolar moment. In the present work,  using data collected from the advanced LIGO interferometers, upper limits of the amplitude of the intrinsic strain tensor and other relevant parameters \cite{Abbott2019} have been predicted. 

\noindent     
\section{Theoretical Calculations} 
\label{Section 5}

    It may be observed from Eq.(8) and Eq.(9) that the integral $\int_{0}^{R_c} \varrho(r) r^6 dr$ is of prime importance for estimating various times scales. This integral can be expressed in a dimensionless form by rewriting it in terms of energy density $\varepsilon(r)=\varrho(r)c^2$ as 

\begin{equation}
I({R_{c}})=\int_{0}^{R_c} \left[\frac{\varepsilon(r)}{\rm MeV fm^{-3}}\right]\left(\frac{r}{\rm km}\right)^6 d\left(\frac{r}{\rm km}\right)
\label{eq35}
\end{equation}

    From Eq.(8) and Eq.(14), the fiducial gravitational radiation timescale $\widetilde{\tau}_{GR}$ can be given by
		
\begin{equation}
\widetilde{\tau}_{GR}=-0.7429\left[\frac{R}{\rm  km}\right]^{9} \left[\frac{1M_{\odot}}{M}\right]^{3}\left[I(R_c)\right]^{-1} ({\rm s})
\label{eq36}
\end{equation}
where $M$ in $M_{\odot}$ and $R$, $r$ are in km.
  
\begin{figure}[ht!]
\vspace{0.0cm}
\eject\centerline{\epsfig{file=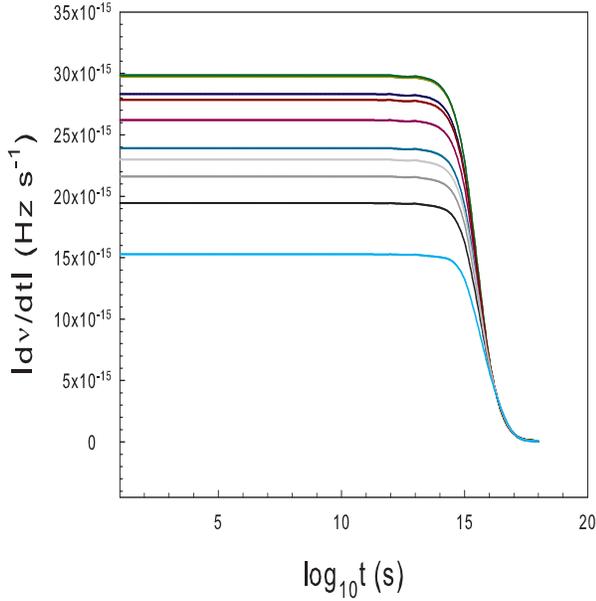,height=8cm,width=7.85cm}}
\caption{Plots of the evolving spin-down rates as a function of time.} 
\label{fig11}
\vspace{0.0cm}
\end{figure} 

    The electron-electron and neutron-neutron scattering fiducial shear viscous timescales $\widetilde{\tau}_{SV}$ may be obtained from Eq.(9), Eqs.(11-13) as 
    
\begin{eqnarray}
\widetilde{\tau}_{ee}=0.1446\times10^8\left[\frac{R}{\rm km}\right]^{3/4}\left[\frac{1M_{\odot}}{M}\right]^{1/4} \left[\frac{\rm km}{R_c}\right]^{6}\nonumber \\
\times\left[\frac{\rm g~cm^{-3}}{\varrho_t}\right]^{1/2}\left[\frac{\rm MeV fm^{-3}}{\varepsilon_t}\right] \left[I(R_c)\right] ({\rm s})
\label{eq37}
\end{eqnarray}
        
\begin{eqnarray}
\widetilde{\tau}_{nn}=19\times10^8\left[\frac{R}{\rm km}\right]^{3/4}\left[\frac{1M_{\odot}}{M}\right]^{1/4}\left[\frac{\rm km}{R_c}\right]^{6}\nonumber \\
\times\left[\frac{\rm g~cm^{-3}}{\varrho_t}\right]^{5/8}\left[\frac{\rm MeV fm^{-3}}{\varepsilon_t}\right] \left[I(R_c)\right] ({\rm s})
\label{eq38}
\end{eqnarray}
where core-crust transition density $\varrho_t$ has been expressed in units of g cm$^{-3}$ while the energy density $\varepsilon_t$ at the transition has been expressed in units of MeV fm$^{-3}$.
 
\noindent
\section{Results and discussion}
\label{Section 6}

    The fiducial timescales versus the gravitational masses of NSs have been plotted in Fig.-6 for the NRAPR EoS. While the timescale for gravitational radiation falls off rapidly, it is observed that the timescales for viscous damping increase almost linearly with increasing mass. From Eq.(16), the dependence of the critical angular velocity $\Omega_c$ on temperature for $(l=2)$ r-mode may be explored by using the fiducial timescales of gravitational radiation and shear viscous drag. The dimensionless ratio $\frac{\Omega_c}{\Omega_0}$ has been shown in Fig.-7 as a function of temperature for distinct NS masses obtained using the NRAPR EoS. These plots define the boundaries of the window of the r-mode instability. The NSs whose angular frequency $>\Omega_c$ lie above the plots and possess unstable r-modes. Hence these NSs reduce their angular frequencies by emitting GWs. Once their angular frequencies reach $\Omega_c$ they reach the region underneath the plots when the resulting r-mode stability causes cessation of GW emissions. In Fig.-7, which depicts the instability windows, the fiducial shear viscous timescale $\widetilde \tau_{ee}$ of Eq.(44) is put in Eq.(16) in place of $\widetilde \tau_{SV}$ in the temperature range T $\leq$ $10^9$ K and $\tau_{nn}$ of Eq.(45) is substituted for $\widetilde \tau_{SV}$ in the temperature range T $>$ $10^9$ K. 

    The frequencies of spinning and the temperatures of the core (either measured or the estimated upper limits) of the observed Low Mass X-ray Binaries (LMXBs) and Millisecond Radio Pulsars (MSRPs) \cite{Ha12} have been collected in Table-II and their positions have been shown in Fig.-8 along with critical frequency versus temperature plots. The Fig.-8 implies that all of the observed NSs lie in the stable region of the r-mode oscillations defined using NRAPR EoS with a rigid crust and have comparatively small amplitudes of r-mode oscillations, which is consistent with the fact that GW emission from these NSs due to r-mode instability have not been observed.		
				
\begin{table}[htbp]
\centering
\caption{The frequencies of spinning and the temperatures of the cores (both the measured values and the estimated upper limits) of the observed \cite{Ha12} Millisecond Radio Pulsars (MSRPs) and Low Mass X-ray Binaries (LMXBs).}
\begin{tabular}{||c|c|c||}
\hline
\hline
 Stellar origin&$ \nu$ (s$^{-1}$)&$ T_{core} (10^8 K) $ \\ 
\hline
 Aql X-1&$550$&$1.08$ \\ 
\hline
 4U 1608-52&$620$&$4.55$ \\ 
\hline
 KS 1731-260&$526$&$0.42$ \\ 
\hline
 MXB 1659-298&$556$&$0.31$ \\ 
\hline
 SAX J1748.9-2021&$442$&$0.89$ \\
\hline 
 IGR 00291+5934&$599$&$0.54$ \\ 
\hline
 SAX J1808.4-3658&$401$&$<0.11$ \\ 
\hline
 XTE J1751-305&$435$&$<0.54$ \\ 
\hline
 XTE J0929-314&$185$&$<0.26$ \\ 
\hline
 XTE J1807-294&$190$&$<0.27$ \\ 
\hline
 XTE J1814-338&$314$&$<0.51$ \\ 
\hline
 EXO 0748-676&$552$&$1.58$ \\ 
\hline  
 HETE J1900.1-2455&$377$&$<0.33$ \\ 
\hline  
 IGR J17191-2821&$294$&$<0.60$ \\ 
\hline  
 IGR J17511-3057&$245$&$<1.10$ \\ 
\hline  
 SAX J1750.8-2900&$601$&$3.38$ \\ 
\hline
 NGC 6440 X-2&$205$&$<0.12$ \\ 
\hline
 Swift J1756-2508&$182$&$<0.78$ \\ 
\hline
 Swift J1749.4-2807&$518$&$<1.61$ \\ 
\hline
 J2124-3358&$203$&$<0.17$ \\ 
\hline 
 J0030+0451&$205$&$<0.70$ \\ 
\hline 
\hline
\end{tabular} 
\label{table4}
\end{table}
\noindent      

\begin{figure}[ht!]
\vspace{0.0cm}
\eject\centerline{\epsfig{file=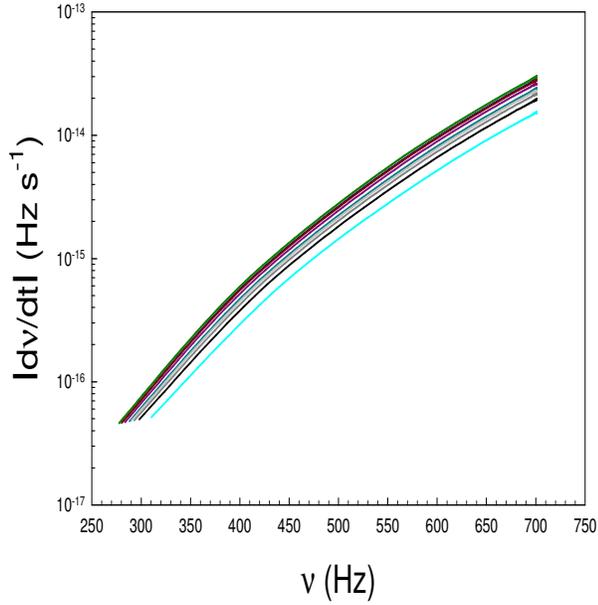,height=8cm,width=7.85cm}}
\caption{Plots of the evolving spin-down rates as a function of frequencies.} 
\label{fig12}
\vspace{0.0cm}
\end{figure}
\vspace{0.0cm}

    For a fixed temperature $T$, $\frac{\Omega_c}{\Omega_0}$ rapidly decreases with increasing mass as seen in Fig.-7. Eq.(18) implies that when $\Omega_c=\Omega_K$, $T=T_c$ and $\Omega_K$ grows with mass leading to a fall in $T_c$. The critical temperature as a function of NS mass has been plotted in Fig.-9. For the calculation of $T_c$, the timescale of shear viscous drag arising from electron-electron scattering has been used. It is observed that the critical temperature $T_c$ decreases rapidly with NS mass.   
 
    It is observed from Fig.-8 and Fig.-9 that $\nu_c$ and $T_c$ decrease with mass causing the r-mode instability window to increase with NS mass. This implies that for the same temperature and EoS, compared to the less massive NSs, more massive configurations are prone to a great extent to the r-mode instability and hence to the emission of GWs. The fact that $\widetilde \tau_{GR}$ is much less than $\widetilde \tau_{ee}$ and $\widetilde \tau_{nn}$ for massive NSs as in Fig.-6, the preceding observation can also be substantiated indirectly. The isolated young massive NSs thus are more prone to GW emissions because of r-mode instability.          		
  
    It is worthwhile to mention here that $\Omega_c$ is sensitive to the density dependence of the symmetry energy and its slope $L$. Moreover, full radius $R$, core radius $R_c$, $I(R_c)$ and $\rho_t$ also depend upon the values of $L$. Therefore, for fixed mass and temperature, the quantity $\Omega_c$ depends upon the above parameters through the relation,
    
\begin{equation}
\Omega_c \sim \frac{R_c^{12/11}}{[I(R_c)]^{4/11}}\varrho_t^{3/11}
\label{eq39}
\end{equation} 

\begin{figure}[ht!]
\vspace{0.0cm}
\eject\centerline{\epsfig{file=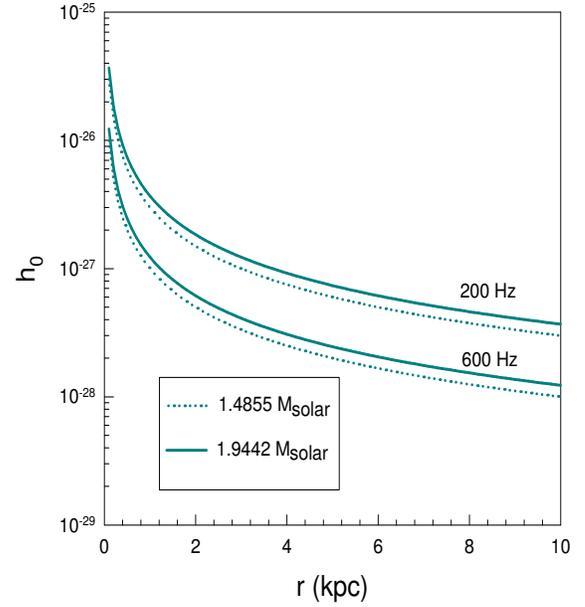,height=8cm,width=7.4cm}}
\caption{Plots displaying the amplitude of the strain tensor ($h_0$) in relation to the distance for two distinct rotational frequencies of neutron stars with varying masses, while the r-mode amplitude estimated from the `thermal equilibrium'.} 
\label{fig13}
\vspace{-0.2cm}
\end{figure}
    
    In our case, for a fixed temperature and NS mass $L$, $\rho_t$ and $R_c$ are constants. The amplitude $\alpha_r$ of the r-mode grows till it attains a saturation value when a NS enters into the region of instability because of mass accretion from its binary companion. At this point a NS emits GWs releasing its energy and angular momentum leading to spin down to reach the stability region. The time evolution of angular velocity of spinning and the rate of spin down can be calculated for a NS from Eq.(22) and Eq.(24), respectively, assuming the ideal condition that the decrease in temperature due to emission of GWs is compensated by the heat produced due to viscous effects, provided T, M, $\alpha_r$ and $\Omega_{in}$ of the star are known. The spin evolution is calculated for different NS masses using the representative values of $\nu_{in}=\frac{\Omega_{in}}{2\pi}=700$ Hz and $\alpha_r=2 \times 10^{-7}$ as used by Moustakidis \cite{Moustakidis2015} and shown in Fig.-10. The rates of spin down are shown for those masses in Fig.-11. The rates of spin down as a function of frequency of spinning are shown in Fig.-12.

    In a previous study \cite{Vidana2012} using different EoS models, the critical frequency $\Omega_c$ of the pulsar 4U 1608-52 as a function of the symmetry energy slope parameter $L$ was plotted using an estimated temperature $\sim 4.55\times 10^8$ K of the core. Like Fig.-6 of \cite{Vidana2012}, using the measured frequency of spinning and the estimated temperature of core, if the mass of 4U 1608-52 is $1.4M_\odot$ then it should be only slightly unstable ($\Omega_c$ is lower than its frequency of spinning), since the radius corresponding to this mass is $\sim 11.87$ kms from the present mass-radius relationship (Fig.-5) and is greater than 11.5 kms. As $L<60$ MeV for NRAPR EoS, the highest mass configuration of 1.9445 M$_\odot$ with a radius of $\sim$9.93 kms is also likely to be in the region of instability. In agreement with \cite{We12}, if one considers the dissipation to arise at the crust-core interface, it would imply that the instability window of the r-mode would be broadened for an isolated NS with a rigid crust.  
    
\begin{figure}[ht!]
\vspace{0.0cm}
\eject\centerline{\epsfig{file=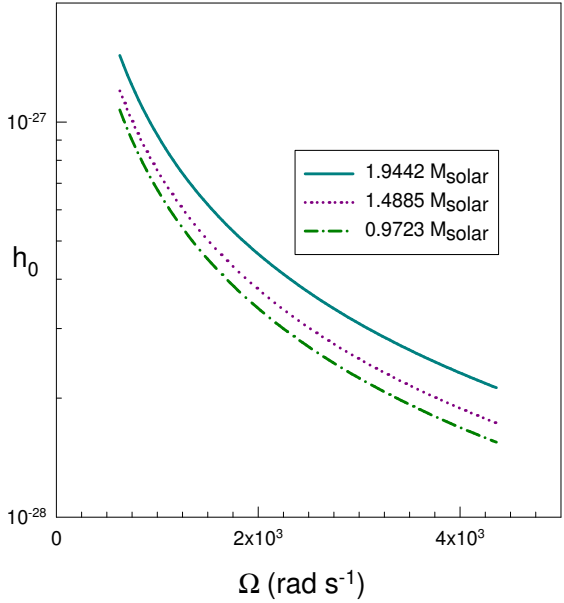,height=8cm,width=7.4cm}}
\caption{Plots displaying the amplitude of the strain\\ tensor ($h_0$) in relation to the angular frequency of the\\ neutron stars with varying masses, while the r-mode\\ amplitude estimated from the `thermal equilibrium'.} 
\label{fig14}
\vspace{0.0cm}
\end{figure}

    It is worthwhile to acknowledge the observations made by Demorest et al. \cite{De10} of the binary millisecond pulsar J1614-2230 that rotates with 3.1 ms period which suggests that its mass lies in the range $1.97\pm0.04$ M$_\odot$. The measurements of radio timing for pulsar PSR J0348+0432 and its companion (white dwarf) have confirmed the mass of the pulsar to be in the range 1.97$-$2.05 $M_\odot$ at the confidence level of 68.27$\% (1\sigma)$ or 1.90$-$2.18 $M_\odot$ at the confidence level of 99.73$\% (3\sigma)$ \cite{An13}. Very recently, the studies for PSR J0740+6620 \cite{Fon21} and for PSR J0952-0607 \cite{Rom22} find masses of 2.08 $\pm$ 0.07 $M_\odot$ and 2.35 $\pm$ 0.17 $M_\odot$, respectively, but only with a $1\sigma$ credibility. Some recent works \cite{Leg21,Ril21} constrain the equatorial radius and mass of PSR J0740+6620 to be 12.39$^{+1.30}_{-0.98}$ km and 2.072$^{+0.067}_{-0.066}$ $M_\odot$ respectively, each reported as the posterior credible interval bounded by the 16$\%$ and 84$\%$ quantiles. All these imply that NRAPR somewhat underestimates the observed NS masses at the $1\sigma$ level measurement credibility but is reasonably good at predicting observations at the $3\sigma$ level.
    
\begin{figure}[ht!]
\vspace{0.0cm}
\eject\centerline{\epsfig{file=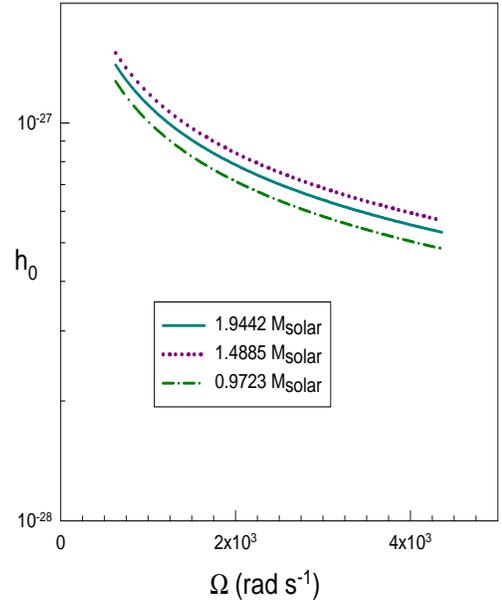,height=8cm,width=6.46cm}}
\caption{Plots displaying the amplitude of the strain tensor ($h_0$) in relation to the angular frequency of neutron stars with varying masses, while the r-mode amplitude estimated from the `spin equilibrium'.} 
\label{fig15}
\vspace{0.0cm}
\end{figure}
    
    The plots of strain tensor amplitude $h_0$ as a function of distance for two different NS masses for two spin frequencies of 200 Hz and 600 Hz are shown in Fig.-\ref{fig13}. The quantity $\alpha_r$ is estimated from 'thermal equilibrium' using the red-shifted effective surface temperature $T_{eff}^\infty = 100$ eV. For same $T_{eff}^\infty$, effective surface temperature $T_{eff}$ for higher mass would be higher which explains the trend resulting from higher $\alpha_r$ and correspondingly higher strain tensor amplitude $h_0$. In Fig.-\ref{fig14}, the strain tensor amplitude $h_0$ is plotted as a function of evolving angular velocity $\Omega$ of the star in units of rad s$^{-1}$ for three different NS masses which arise from GW emissions. In Fig.-\ref{fig15}, $\alpha_r$ is estimated from 'spin equilibrium' using Eq.(\ref{eq25}) where the experimental data used were from the source IGR J00291 of Table-II of Ref.\cite{Mahmoodifar2013}. Entire calculations have been performed presuming the range of angular momentum of the star such that the CFS two-stream instability \cite{Chandrasekhar1970} applies enhancing the r-mode perturbation as the GW emission pumps into the perturbation instead of damping it. Thus the NS, spinning at a particular angular velocity emits GW because of the perturbation and causing loss of its angular momentum. As a consequence while the the star's angular velocity continues to reduce till it crosses the critical limit ($\Omega_c$) of CFS two-stream instability, the intensity of GW emission increases.  
    		
\noindent
\section{Summary and conclusions}
\label{Section 7}

    The current study explores the instability due to the r-mode oscillations for slowly rotating neutron stars having a solid crust, by utilizing an equation of state derived from the Skyrme effective nucleon-nucleon interaction with the NRAPR parametrization. The fiducial gravitational radiation timescale and the timescales for shear viscosity drag have been calculated in the framework of Skyrme-NRAPR interaction for a wide range of NS masses. Using the present EoS, it is seen that while the viscous damping timescales increase more or less linearly with increasing NS mass, the GW emission timescales decrease rapidly. Further studies have been conducted on the variation of the critical angular frequency with temperature for diverse NS masses. As observed in Fig.-8, the core temperatures and spin frequencies of the observed Millisecond Radio Pulsars \cite{Ha12} and Low Mass X-ray Binaries always lie below the region of the r-mode instability. This fact implies that for the NSs spinning faster than their corresponding critical frequencies have unstable r-modes leading to the emission of GWs. In the present study it is observed that with increasing mass the temperature and critical frequency decrease. It may therefore be concluded that hot massive NSs are more vulnerable to instability caused by r-mode through GW radiation. The rates of spin down and the time evolution of angular frequency of NSs due to the r-mode instability have been calculated as well. We highlight the dependence of the critical angular frequency on the EoS by examining the radius and the slope parameter $L$ of symmetry energy. The region of the r-mode instability extends to the lower values of $L$ if the r-mode dissipation arising out of shear viscous drag proceeds along the layer separating the core from the crust at the core-crust interface. 
    
\begin{acknowledgements}

    One of the authors (DNB) acknowledges support from Science and Engineering Research Board, Department of Science and Technology, Government of India, through Grant No. CRG/2021/007333.

\end{acknowledgements}		

%\clearpage

\end{document}